# DISTRIBUTED DENIAL OF SERVICE (DDoS) ATTACKS DETECTION MECHANISM

Saravanan kumarasamy[1] and Dr.R.Asokan[2]

[1]Department of Computer Science and Engineering, Erode Sengunthar Engineering College, Thudupathi, Erode -57, India
saravanankumarasamy@gmail.com
[2]Professor and head, Department of IT, Kongu Engineering College, Erode

## ABSTRACT

*Pushback is a mechanism for defending against Distributed Denial-of-Service (DDoS) attacks. DDoS attacks are treated as a congestion-control problem, but because most such congestion is caused by malicious hosts not obeying traditional end-to-end congestion control, the problem must be handled by the routers. Functionality is added to each router to detect and preferentially drop packets that probably belong to an attack. Upstream routers are also notified to drop such packets in order that the router's resources be used to route legitimate traffic hence term pushback. Client puzzles have been advocated as a promising countermeasure to DoS attacks in the recent years. In order to identify the attackers, the victim server issues a puzzle to the client that sent the traffic. When the client is able to solve the puzzle, it is assumed to be authentic and the traffic from it is allowed into the server. If the victim suspects that the puzzles are solved by most of the clients, it increases the complexity of the puzzles. This puzzle solving technique allows the traversal of the attack traffic throughout the intermediate routers before reaching the destination. In order to attain the advantages of both pushback and puzzle solving techniques, a hybrid scheme called Router based Pushback technique, which involves both the techniques to solve the problem of DDoS attacks is proposed. In this proposal, the puzzle solving mechanism is pushed back to the core routers rather than having at the victim. The router based client puzzle mechanism checks the host system whether it is legitimate or not by providing a puzzle to be solved by the suspected host.*

## 1. INTRODUCTION

Denial of Service (DoS) attack is executed to determine a specific category of information warfare where a malicious user blocks legitimate users from accessing network services by exhausting the resources of the victim system. Without hacking the password files or stealing sensitive data, a DoS attacker creates network congestion by generating a large volume of traffic in the area of the targeting system. The size of the caused overload is enough to prevent any packet from reaching its destination. Normally, a TCP connection is established through a three-way handshake. A client initiates a connection by sending a SYN packet to the server. The server acknowledges the request by sending a SYN ACK packet back to the client and allocating space for the connection in a buffer. The client then replies with an ACK packet, and the connection is completely established. In the TCP SYN flood attack, an attacker initiates many connections in which the SYN and SYN ACK packets are exchanged as usual, but the final ACK message is never sent to the server. Thus the connection is never completely established, and the server is left with buffer space allocated for all the incomplete connections. If the attack succeeds, the server fills up its buffer with incomplete connections, leaving no space for non-malicious connection requests, and thus preventing the server from establishing DoS attack can be characterized as an attack with the purpose of preventing legitimate users from using a victim computing system or network resource. A Distributed Denial of Service (DDoS) attack is a large-scale, coordinated attack on the availability of services of a victim system or network resource, launched indirectly through many compromised computers on the Internet. The services under attack are those of the "primary victim", while the compromised systems used to launch the attack are often called the "secondary victims." The use of secondary victims in performing a DDoS attack provides the attacker with the ability to wage a much larger and more disruptive attack, while making it more difficult to track down the original attacker connections with legitimate (non-malicious) clients.

The DDoS attacks are among the hardest security problems today to detect, defend and trace because of the limitation of current network components, multiplicity of attack methods and invisibility of the operators to host sites. While former simple forms of DoS can be defeated by tight security policies and active measures like firewalls and vendor-specialized patches, there is no completely effective defense from DDoS attacks. In recent experiments, SYN attack, one of the most well known DDoS attacks on commercial platforms, showed that an attack rate of only 500 SYN packets per second is enough to overwhelm a server. In detail, 38% of uniform random attack events and 46% of all attack events had an estimated rate of 500 packets per second or higher. The same experiments showed that even a specialized firewall, which is designed to resist SYN floods, becomes futile under a flood of 14,000 packets per second.

Difficulties in providing the powerful mechanism against DDoS attacks is finding out the traffic pattern which causes to congestion due to an attack, and conforming the suspected hosts whether the host is really an attacker or a normal host. There were many challenges against DDoS attack defense mechanism; those are to say a good mechanism it must have to identify the attack before the first successful attack. After the first successful attack the victim lost its control capability against attack.

## 2. EXISTING SYSTEM

Inferring of Internet Denial-of-Service Activity is extremely difficult for even system administrators because the current Internet protocol does not require a mechanism for the current packets to be pre-verified before leaving from a source network, during traversing through inter-networks and finally to be authenticated after arriving at any machines of the destination network. This anonymity is the most beauty of the Internet in spite of leaving unfavourable security issues.

### 2.1. TCP SYN Flood

The TCP SYN flood is one of the most dangerous forms of the DDoS attacks. When two sides want to establish a TCP connection, the system, which asks for the connection (client), has to send initially a "SYN" message to the other system (server) in order to notify its intention. When the server receives the "SYN" message, it reserves some of its resources for the expected connection and sends a "SYN-ACK" message back to the client. The reception of "SYN-ACK" message from the client triggers the transmission of a new message "ACK" which is sent to the server in order to complete the last stage of a three-step handshake protocol. After reception of the last message "ACK" from the server, the connection is successfully established and the two peers are able to start exchanging their data. If the server does not receive the "ACK" message from the other side then it discards the partially established connection and releases the set of resources reserved for the specific attempt. This three way handshake is shown in figure 1. In

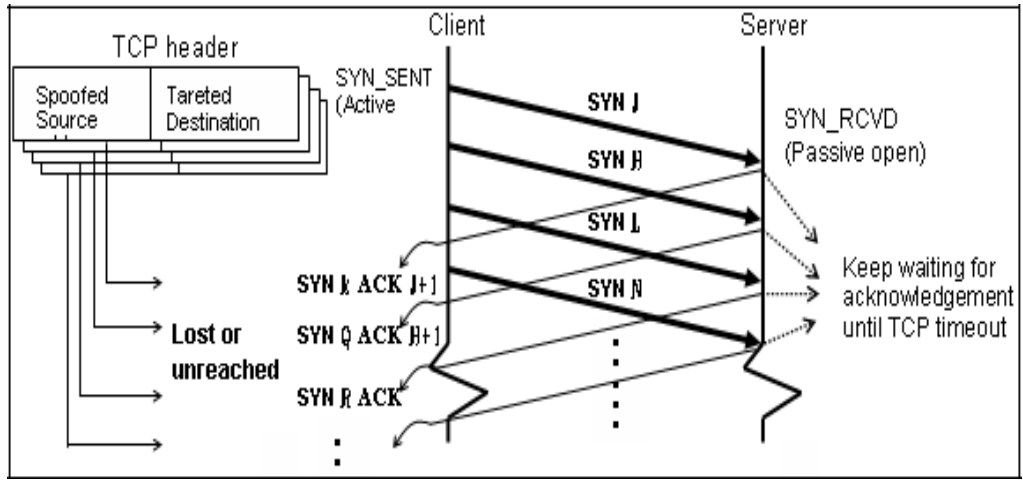

Figure 1 Three way handshake of TCP

Order to keep a track of the requests, the server builds a backlog queue into its system memory. This queue maintains a limited number of half-open connections per port. Once the backlog queue limit is reached, the server discards every new connection request from the clients.

## 2.2 Ingress /Egress Filtering

Most of DoS/DDoS attacks use forged or spoofed source IP addresses in order to hide the attacker's originality and also indirectly generate the massive traffic from the intermediary network to the target machine. As a result, a machine, that the spoofed address is belonging to, is also a victim of the attack. A packet leaving to Internet and arriving from Internet must have a source address originating from an interior network. By blocking packets with non-local source IP address from leaving a interior network, DoS/DDoS attacker's source address spoofing become impossible. However, even though this scheme is most feasible in customer network, the universal deployment is not likely to be accomplished because of administrative burden, potential router overhead and complications with existing services that depend on source address spoofing. Moreover, if an interior network is quite large, or each sub-network does not have the address filtering capability, the attackers could still forge addresses from hundreds of thousands of hosts within a valid interior network.

## 2.3 IP Trace back

Most existing traceback techniques start from the router closest to the victim and interactively test its upstream links until they determine which one is used to carry the attacker's traffic.

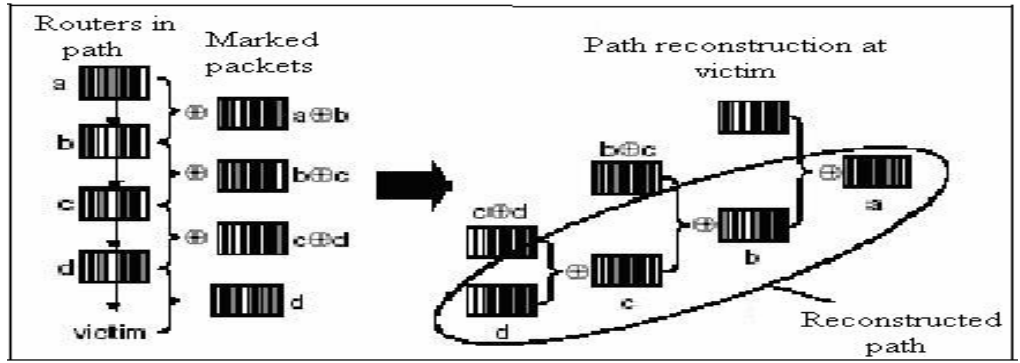

Figure 2 IP Traceback

Ideally this procedure is repeated recursively on the upstream router until the source of traffic is reached. This technique has a critical assumption that an attacker will be remained while tracing mechanism is completed. To avoid the overhead of trace back, Burch and Cheswick proposed the possibility of tracing the flooding attacks by "marking" the packets, either probabilistically or deterministically, with the address of the routers they traverse. Therefore, the victim can use this marking information to trace an attacker back to its source. Figure 2 gives an example of path reconstruction proposed by Savage et.al.

## 2.4 Smurf

The two main components to the Smurf Denial-of-Service attack are the use of forged ICMP echo request packets and the direction of packets to IP broadcast addresses. The Internet Control Message Protocol (ICMP) is used to handle the errors and exchange control the messages. Also, the ICMP can be used to determine whether a machine on the Internet is responding properly or has connection problems. To do this, an ICMP echo request packet is sent to a machine with a return address that the contacted machine would return an ICMP echo reply packet when receiving the ICMP echo request packet. This process can be seen in fig. 3. The most common implementation of this process is the "ping" command, which included with many operating systems and network software packages.

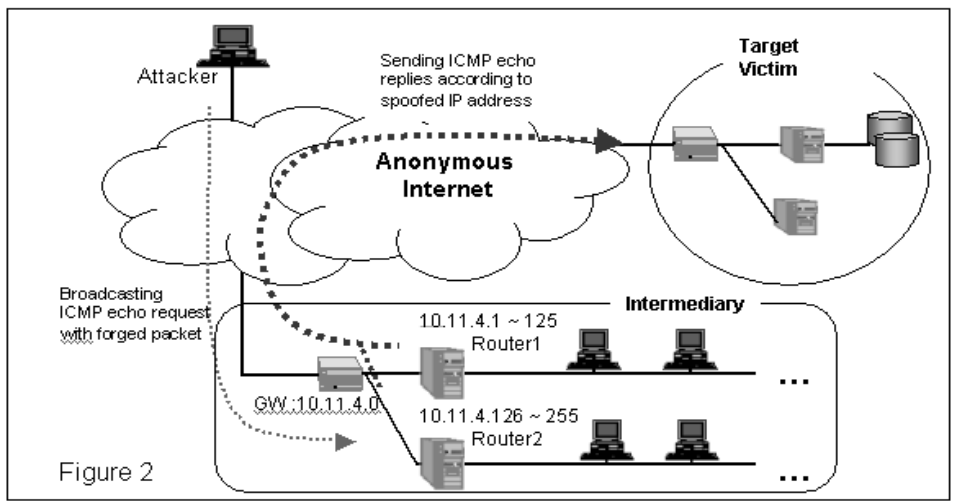

Figure 3 SMURF attack scenario

ICMP can also be a valuable tool in diagnosing host or network problem, because it conveys status and error information, including the notification of the network congestion and of other network transport problems.

## 2.5 UDP Flood

This is the second most popular DDoS attack method after TCP SYN flood. The basic idea in the UDP Flood attacks is to exploit UDP services, which are known to reply to the packets. The hacker is armed with a list of broadcast addresses, to which he sends spoofed UDP packets. These packets are sent to random and changing ports of the unsuspected target location. In most of the cases the packets are directed to the echo port 7 (echoes any character it receives in an attempt to test network programs) on the target machines.

However, there are attacks in which the malicious user sends packets to the chargen port. The chargen port is a port, which is used for testing purposes and generates a series of characters for each packet it receives. By connecting a host's chargen service to the echo service on the same or another machine, all affected machines can be effectively taken out of service as an excessively high number of packets are going to be produced. In addition, if two or more hosts are so connected, the intervening network can also become congested and deny service to all hosts whose traffic traverses that network.

It is obvious from the previous analysis that the result from a UDP flood attack is the creation of a nonstop flood of useless data passes between two or more systems. The target host returns ICMP port unreachable messages as a response to each spoofed UDP packets and then slows down because it becomes more and more busy in processing the forged IP addresses. This "loop" is responsible for the overload of the network and the total exhaust of the available bandwidth. Victims of this massive amount of traffic can be also individual systems which can lose connectivity to the Internet and in some cases, crash.

## 2.6 Traffic Shaping

A number of routers in the market today have features to limit the amount of bandwidth that some type of traffic can consume. This is sometimes referred to as "traffic shaping". This can be used in a proactive way if the traffic behaviour of the network is already known. It can also be used in a reactive way by crafting an access rule that would match some of the network traffic using by the DDoS attack. For example if the attack is employing ICMP packets or TCP SYN packets you could configure the system to specifically limit the bandwidth those types of packets. This would allow some of these packets that may belong to legitimate network flow to go through.

Because of the avalanche affect of the DDoS attacks for this option to be effective, it must be deployed as depth into the network as possible (closer to the source of the attack packets). Would the ISPs may be required to implement these filters in their routers. This would not be possible for many organizations for a number of reasons. Furthermore, DDoS attack tools can generate random packets such as that matching them with a set of access list. Rules can become difficult unless by using negative space while defining normal traffic pattern and assuming everything else is DDoS traffic.

## 2.7 Traffic Analysis

A number of researchers in the academic fields have proposed different approaches to analyze the traffic patterns in order to infer the attacking packets and its characteristics. Most of methods are detecting the pattern of illegitimate packets or their source using probabilistic and statistic analysis. A critical point of those researches is that a large scale attacks can readily be identified by observing very abrupt changes in the network traffics and most of packets would have a certain type of pattern so that we can classify them according to each pattern.

For example, at first, randomly collect sample, classify the collected packets, and then normalize the data or build a temporary DB. Second, using a specific algorithm and modeling, find a pattern for bad-will packets from the data sampling. Most of differentiated methods are developed in second phase such as using Time series, Data Mining, Probabilistic Modeling and more complicated mathematical models.

However, even though those methods can provide quite reasonable solutions to detect bad-will packets, we cannot be fully confident that every attacking packet can be detected or only illegitimate packets are likely to be detected since these mechanisms are relying on probabilistic model

## 3. PROPOSED SYSTEM

Establishing a defense mechanism which provides a powerful security against resource consumption attacks namely denial of service attacks. Before the successful attack the attacker's traffic consumes much network resources which lead to congestion over the network. This mechanism is a hybrid model consists of pushback mechanism with client puzzles. With this mechanism the attack causing traffic pattern can be easily identified and defense against the attack will be done successfully.

The proposed system can be easily implemented onto the ISP networks.

1. Implanting the intelligent router over the ISP network
2. It pushback the suspected hosts to upstream router
3. The upstream router issues the puzzle to suspected host to validate
4. Edge router blocks the attack traffic

### 3.1 Concept

The intelligent router over the ISP network identifies the attacker hosts and pushes back the host addresses to the upstream router. The upstream router upon receiving the pushback request with the suspected host addresses it gives a puzzle to suspected host, the suspected host have to solve the puzzle and send back the puzzle answer to the router issued puzzle. Normally resource consumption attacks are not made by human requests, it makes resource consumption attack, and attackers write a program to continuously keep on requesting the server. So if it is an attacker host it can't solve the puzzle, then it is conformed as an attacker hosts. The router issued the puzzle conforms the suspected hosts as attackers and asks the edge router to block all the traffic from attacker hosts.

### 3.2 Phases in proposed method

The proposed work is a hybrid model for providing the defense against DoS/DDoS attacks. A router based client puzzles to suppress the attack traffic at the edge router itself is introduced. Intelligent routers have the responsibility of authenticating the host requests and allocating the network resources. This router based model is integrated with the pushback method. With this a very powerful defense against both the DoS/DDoS attacks can be provided.

The proposed defense system is divided into four phases. The first phase is the Network Model, second phase is Good and Attack Traffic Generation, third phase is Pushback Mechanism and fourth phase is Client Puzzles.

## 4. IMPLEMENTATION

To simulate the proposed system, have to create the network. This network consists of LAN's connected with ISP's. This Network layout was created in Network Simulator. Figure 4 shows the network layout structure .In the below figure L1, L2 and L3 represents LANS Connected with ISP's. In network simulator all the nodes will have same functionality and capabilities, to implement the proposed system in the above network; separate functionality in the nodes is implemented.

Nodes with square shape have the router functionality; nodes with the hexagonal shape are the active hosts on the network. Node-1 and Node-19 are used as legitimate users, Node-5 and Node-20 are used as attacker hosts, Node-24 is the intelligent router it have the capabilities to check the traffic pattern and identify the attack causing events.

### 4.1 Attack Traffic Generation Phase

Once the network layout is generated, traffic source is to be added to the both legitimate user and attackers. The attacker traffic source is formed in such a way that it won't obey the traffic rules and try to utilize the full network resources. If any network congestion takes place, it won't slowdown its traffic generation rate and it keeps on sending packets. The legitimate user follows the traffic rules and if any congestion takes place, it slowdowns its sending rate. After the congestion, it checks whether the route is free or not. If the route is still congested, it waits for some more time the network traffic to clear.

The legitimate user requests the server and the server opens a connection for it and the legitimate user uses the connection like three way hand shake. Attacker requests the server then the server opens the connections, but the attacker won't use the connection and keeps on requesting many connections. Due to this the server connections are wasted and sever is not able to serve legitimate user.

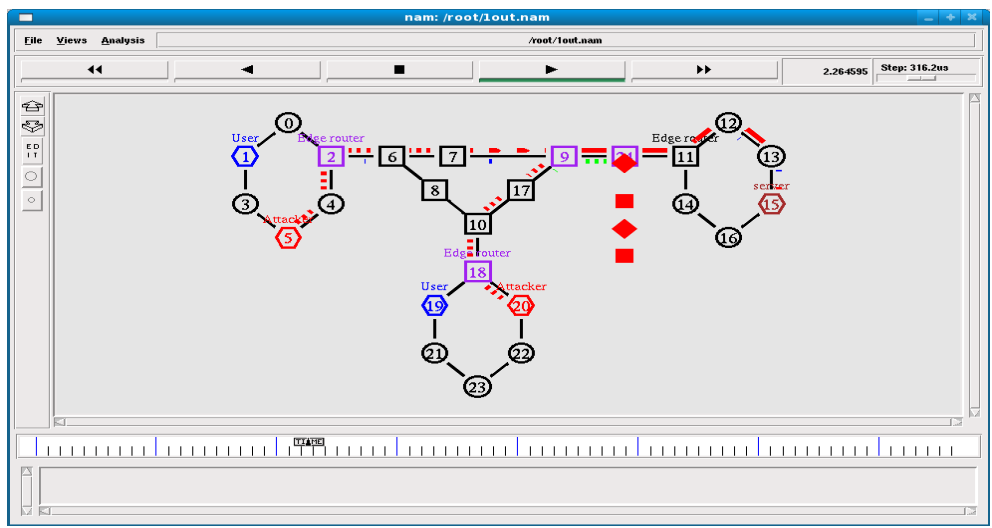

Figure 4 Network Model

## 4.2 Pushback Mechanism Phase

This is the more important phase, where the intelligent router initiates pushback mechanism and there to client puzzles. In this phase, a router in the ISP network is selected as intelligent router and a powerful mechanism is implemented on this router.

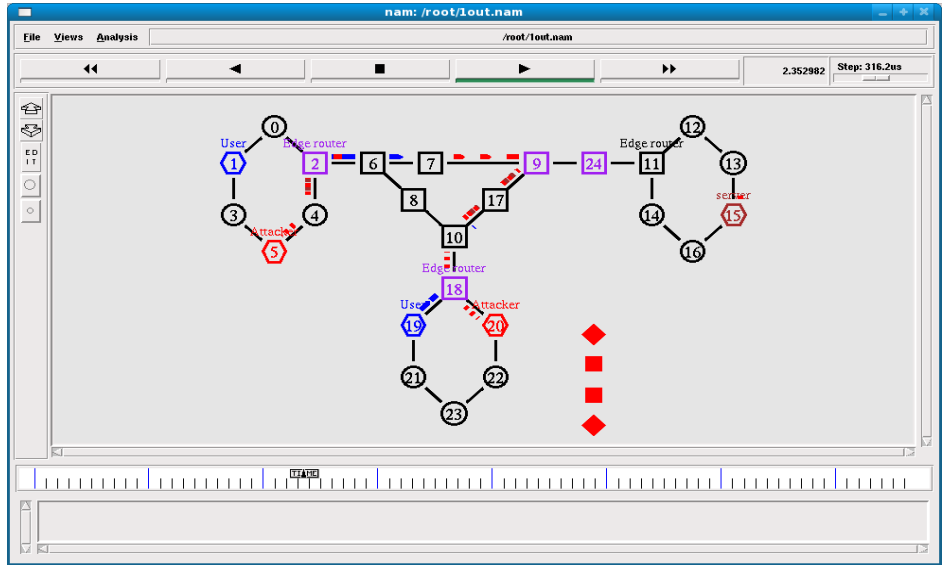

Figure 5 Puzzle Validation

The implementation of pushback mechanism is as follows. Making congestion Signature, Matching the traffic pattern, updating Congestion Signature and Initiation of Pushback mechanism. With the proposed method, attack traffic over the network is greatly reduced. The proposed system effectively provides defense against DoS/DDoS attacks.

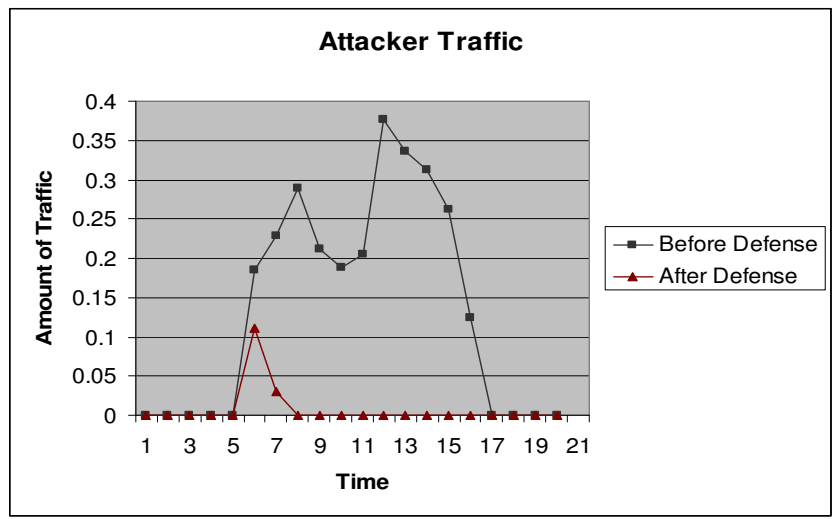

Figure 6 Attacker Traffic

Figure 4 show the effect of defense mechanism and decrease in the attacker traffic over the network. This method identifies the attacker hosts before the attack. Figure 6 shows the amount of attacker traffic over the network before the proposed method and the amount of attacker traffic after the defense mechanism. In this method suspected host is validated with client

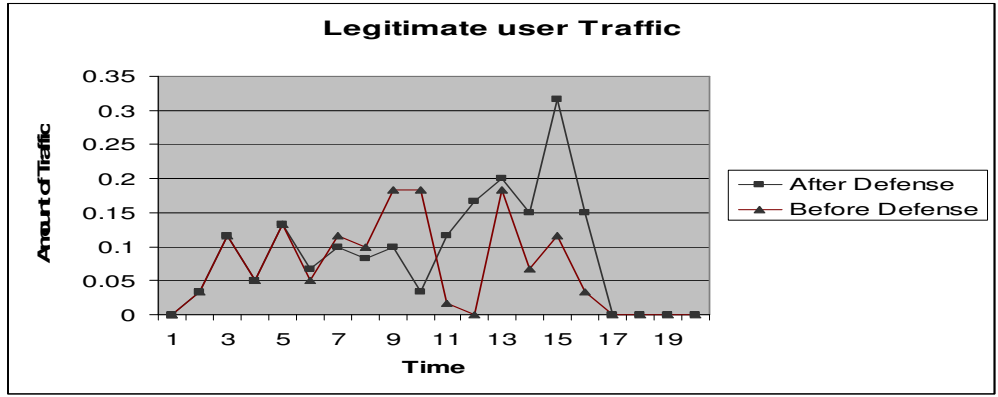

Figure 7 Legitimate User Traffic

Puzzles and after the conformation attacker traffic is dropped at the edge routers. Figure 7 shows the legitimate user traffic strength over the network before and after the proposed system implementation. The attacker traffic consumes much network resources before it reaches the victim (target) thus leading to congestion. With the proposed system, such problems are rectified as the congestion signature can be adjusted and updated in order to find these abnormalities over the network.

## 5. CONCLUSIONS

This proposed method provides the strong defense against the malicious hosts in the network, and it easily identifies the attacker hosts by their traffic nature and blocks all the traffic from the attacker hosts. Client puzzles gives the advantage to validate the suspected hosts in order to conform whether the suspected hosts from an attacker or from a legitimate user. Pushback helps to outsource the client puzzle work load to upstream router, which helps to decrease the processing work load on intelligent router. Using the proposed work, the attacker traffic is effectively blocked at the edge routers and hence the denial of service causing attacks can be identified in advance and offended successfully.

## Authors

**Mr.K.Saravanan** received the M.E degree 2008 in computer science from Dr.MCET, Anna University, Chennai, India. He is currently working as a Lecturer at the Faculty of Engineering, Erode Sengunthar Engineering College, Erode, Tamilnadu. He has published 4 paper in International Journal, 07 papers in National Conference and 02 papers in International Conference..His current research interests are information security, computer communications and DDoS Attacks. He is currently pursuing Ph.D. under Anna University of Technology, Coimbatore.

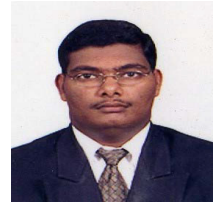

**Dr.R.Asokan** received the Ph.D degree in Information and Communication Engineering from Anna University, Chennai. He has 21 years of teaching experience. His areas of interest include mobile networks, Adhoc Networks and Network Security. At present he is working as Professor and Head in Department of IT at Kongu Engineering College, Perundurai, Erode.

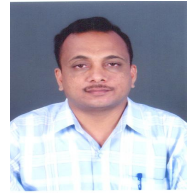